\input harvmac.tex
\pretolerance=10000

\def\phi{\varphi}

\Title{hep-th/0209204, HWS-200203}
{\vbox{\centerline{A Mechanism for Charge Quantization}}}


\centerline{Theodore J. Allen\footnote{$^\dagger$}{tjallen@hws.edu} }
\medskip\centerline{Department of Physics, Eaton Hall}
\centerline{Hobart and William Smith Colleges}
\centerline{Geneva, NY \ 14456 USA}
\vskip .2in
\centerline{Costas J. Efthimiou\footnote{*}{costas@physics.ucf.edu} }
\medskip\centerline{Department of Physics and Astronomy}
\centerline{University of Central Florida}
\centerline{Orlando, FL 32816 USA}
\vskip.2in
\centerline{Donald
Spector\footnote{$^\sharp$}{spector@hws.edu} }
\medskip\centerline{Department of Physics, Eaton Hall}
\centerline{Hobart and William Smith Colleges}
\centerline{Geneva, NY \ 14456 USA}

\vskip .3in
We analyze a potential that produces background charges which are
automatically quantized. This introduces a new mechanism for charge
quantization, although so far it has only been implemented for background
charges.  We show that this same mechanism can also lead to an alternative
means of hiding extra dimensions that is analogous to the Kaluza-Klein
approach.

\Date{09/2002}

\newsec{Introduction}

The quantization of electric charges has been a puzzle since it was
discovered by Millikan \ref\millikan{R.A. Millikan, Phys.\ Rev.\ 2 (1913)
109.} nearly a hundred years ago.  There have been two conventional
explanations of this phenomenon.  The first, due to Dirac \ref\dirac{P.A.M.
Dirac, 
Proc.\ R.
Soc.\ A 133 (1931) 60.}, considered the effect of magnetic monopoles. Dirac
showed that, in the presence of a magnetic monopole, only quantized
electric charges are allowed.\foot{A generalized condition follows for
dyons \ref\zwanziger{ J. Schwinger, Phys.\ Rev.\ 144 (1966) 1087\semi
J. Schwinger, Phys.\ Rev.\ 144 (1968) 1536\semi D. Zwanziger, Phys.\ Rev.\ 176
(1968) 1480\semi D. Zwanziger, Phys.\ Rev.\ 176 (1968) 1489.  }.}  Thus,
should a magnetic monopole be discovered, quantization of electric charge
would, of necessity, follow.  The alternative explanation is one based in
gauge theories.  If the $U(1)$ gauge group of electromagnetism is embedded
in a non-abelian gauge group, then charge quantization is automatic, for
group theoretic reasons \ref\unification{H. Georgi and S. Glashow,
Phys.\ Rev.\ Lett.\ 32 (1974) 438.}.  The commutation relations for the
non-abelian group impose non-abelian charge quantization, and thus the
embedding of the $U(1)$ group in the non-abelian group implies quantization
for the electric charge.

We will here present a theory that exhibits a quantized electric
charge, but in a very different way.  The theory in question will be
an example from non-relativistic quantum mechanics, and give us an
entirely different way to think about the origins of charge
quantization by naturally producing a situation in which there are
background charges, and for which these background charges are quantized.
One of the bonuses of this method is that it also gives us a novel way
to think about the possibility of higher dimensions, and gives us a
way in which such dimensions can appear small without being explicitly
compactified.

In some sense, the structure we identify is more like that typically found
when there are topological charges.  We find discrete sectors of the
theory, sectors which can be understood as being related by the addition of
quantized potential terms.  The difference is that these sectors do not
arise for topological reasons in the case that we study here.

We should note that, at
present, the potential function that we examine in this paper is
constructed in
very much an {\it ad hoc} way.  We do not yet have a natural mechanism for
causing this potential (or one like it) to appear, or
incorporate it in such a way as to lead to quantization of dynamical, as
opposed to background, charges.  However, what we do have, even
with these caveats, is a new conceptual approach to the
question of what can enforce charge quantization or lead to dimensional
compactification.  While we would of course have preferred already to have
developed the full application of this idea, we recognize that often
such work
must be developed in stages.  It is in this light that we therefore present
this first stage here, presenting both the concept and its initial
implementation, while we reserve for future work the further technical
developments that will extend the applicability of these ideas.

\newsec{The Digamma Potential}

The function $\Gamma(x)$ is well-known, of course; it is the analytic
function that generalizes the factorial, with the relation holding
for all arguments
\eqn\gammaid{\Gamma(x+1) = x \Gamma(x)~~~.}
The digamma function is then defined as
\eqn\psidef{\psi(x) = {d\over dx}\ln(\Gamma(x)) = 
{\Gamma^\prime(x)\over\Gamma(x)}~~~.}
This function satisfies the identity
\eqn\psiid{\psi(x+1) = \psi(x) + 1/x~~~.}
We warn the reader not to confuse the digamma function with the
quantum mechanical wavefunction, as both are typically denoted with
the same Greek letter.  We will reserve for ``$\psi$'' for the digamma
function, and will not need any particular symbol for the
wavefunction.

Suppose, now, that we consider the non-relativistic quantum mechanical
Hamiltonian which has the digamma function as its potential,
\eqn\hampsi{H = -{\hbar^2\over 2m}{d^2\over dx^2} + \psi(x)~~~,}
where the coordinate $x$ labels the real axis.
The potential has poles at every non-negative integer. Note that at
large positive
$x$,
$\psi(x)$ behaves asymptotically as $\ln(x)$ (this can be seen from
the
Stirling approximation), while at each negative integer, the function
diverges to $-\infty$ on the right and to $+\infty$ on the left.

Because of the infinities, the behavior of the wavefunction 
at these  negative
integer values must be controlled.  One way to do this would be
to specify put boundary conditions on the wavefunction at these
points, so as to produce a self-adjoint extension.  Our analysis of this
problem indicates that one can do so in a way that imposes the condition 
that at these points the wavefunction vanishes, while its derivative need
not be continuous.  

However, we can also adopt a simpler
approach that achieves the same results.  All the properties of the potential
that will be necessary in our analysis can be preserved if we add a
periodic term to the digamma function potential.  We therefore can
consider the
generalized potential
\eqn\genU{\tilde U(x) = \psi(x) + {1\over \sin^2{\pi x}}~~~.}
This potential still satisfies \psiid , but at the same time causes the
potential to diverge to a positive infinite value at every negative integer,
whether one approaches from the left or the right, as there is now a double
pole at each of these points.  We thus imagine that we have added such a term
to the potential. Note that the
potential we have added is, when considered by itself, exactly soluble, and
related by shape invariance \ref\shape{F. Cooper, A. Khare, and U. Sukhatme,
Phys.\ Rep.\ 251 (1995) 267.} to the infinite square well.  This guarantees
that the boundary conditions in the presence of this new term are such that
the wavefunction vanishes whenever $x$ is a negative integer, with the
wavefunction generally being discontinuous at these points, as the new term
dominates the digamma function at these locations, and hence determines the
behavior of the wavefunction. 

Whichever approach one takes -- either constructing self-adjoint
extensions or adding the extra term to the potential -- will have the same
effects, and will lead to the conclusions presented in this paper.  For the
sake of simplicity of presentation, we will suppress the addition of the
$1/\sin^2(\pi x)$ term, as it does not affect the mechanism behind the
key results we obtain, but the reader should be aware that such a term
can, 
if one chooses the latter scheme, be added throughout the analysis of
this paper, although we will our frame our results without reference to a
specific choice between these two options.

Each interval
$-(m+1) < x <-m$, where $m$ is a non-negative integer, we will term a 
sector.
Since the wavefunction vanishes at the endpoints of each sector, and
the 
derivatives
need not be continuous at these points, we see that wavefunctions
which 
are non-zero in
only one sector can be perfectly good eigenstates of the Hamiltonian,
and
so we can choose to focus on such functions as the eigenfunctions of
interest.  One then solves the Schr\"odinger equation in each sector, 
finding
the corresponding wavefunctions.  Due to the boundary conditions
imposed by either of our techniques, it is acceptable to consider
this  theory exclusively
on the negative real axis, which for simplicity we will do here 
(although we will see later that we can remove this limitation). Thus the
eigenfunctions for the different sectors, when put together, form a 
complete basis
on the space in question.

What would a theory on this half-line look like?  As we discussed
above, 
one
way to think of it would be as a set of separate sectors.  However,
this
theory, as we will show shortly, is equivalent to another theory: one 
with
only a single spatial sector, but with different background charges.
The important feature here is that the background charges must be 
quantized.

Let us refer to the region in the interval $-(m+1)<x<-m$ as the
$m^{th}$ 
{\it sector}
of the theory.  Then when we are studying the wavefunctions confined
to 
this sector, we need
simply solve the Schr\"odinger equation with the potential $V_m(x) = 
\psi(x)$ in the
$m^{th}$ sector, and with barriers preventing penetration into the 
adjacent sectors.

We can, however, label the sectors using shifted coordinates, so that
in 
each sector,
we restrict the spatial variable $x$ to the interval $-1<x<0$.  Then
in 
each sector, we must
solve the Schr\"odinger equation with a suitable potential.  We denote
the potential in the
$m^{th}$ sector as $U_m(x)$.  Then we have
\eqn\uzeroexample{U_0(x) = \psi(x)~~~,}
\eqn\uoneexample{U_1(x) = \psi(x-1) = \psi(x) - {1\over 
x-1}=U_0(x)-{1\over x-1}~~~,}
\eqn\utwoexample{U_2(x) = \psi(x-2) = \psi(x-1)-{1\over x-2} =U_0(x) -
{1\over x-2}
    - {1\over x-1}~~~,}
and, in general,
\eqn\ukexample{U_k(x) = U_0(x) - {1\over x-1}-{1\over x-2} - \cdots 
-{1\over x-k}~~~.}

Consequently, we can think of the theory in sector $m-1$ as equivalent
to the theory in sector $m$ provided that an additional charge, with
its
associated ``Coulombic'' potential\foot{We put the term {\it
Coulombic} 
in quotation
marks, as it is only in three spatial dimensions that the Coulomb 
potential
goes as $1/r$.} having been added
to the theory.  The coefficient of the attendant $1/(x-m)$ term
is a charge of the theory, as
it determines
the strength of the force on the dynamical particle.

Iterating this process as we have in \ukexample, we see that every 
sector is equivalent to the $0^{th}$ sector with
a series of additional charges added to the theory through the
background potential, with each of these additional charges
having strength $-1$.  Thus, the
$m^{th}$ sector can be understood as equivalent to the $0^{th}$ sector
in the presence of
$m$ background charges of charge $-1$.  (These charges also
appear 
at discrete locations.)
This fixed charge strength tells us, then, that these background
charges are quantized.

Thus we have a theory which, although initially formulated on the 
semi-infinite
negative real-axis, should actually be understood as being confined to
a 
single finite
interval, but in the presence of various background charges, with
these 
charges
quantized.  This is the crux of our mechanism: that the potential
itself automatically generates a quantization of charge.
As a consequence, we have a distinct and novel mechanism for
explaining 
the appearance of
charge quantization.  Even though the present model cannot by itself
be 
a realistic model
of nature, the mechanism itself may well prove fruitful when developed
in
more elaborated settings.

We note, too, that there are natural ways to generalize this
phenomenon, 
ways
that are already straightforward to characterize.
For example, as we have already discussed,
one can add any periodic potential to
this system in the original definition of the Hamiltonian \hampsi\ without
altering our  analysis;
thus we are not restricted specifically to the unmodified digamma 
function
potential.  Indeed, this idea underlay our addition in \genU\ of a
$1/\sin^2(\pi x)$ term in lieu of constructing a self-adjoint extension, but
one can of course also add other periodic potentials that are 
everywhere finite, and still maintain the mechanism for the appearance of
quantized background charges.\foot{In fact, now that we have presented the
main analysis, one sees that the division into distinct sectors with quantized
background charges will hold for all $x$, not just the negative half of the
$x$-axis, when we add the $1/\sin^2(\pi x)$ term to the potential.}
One can
also easily imagine extending this to higher dimensions, using, for 
example, a potential in
three-dimensions such as
$U(x,y,z) = \psi(x)\psi(y)\psi(z)$.  This gives rises to a theory
which 
can
be viewed as consisting of various cubical sectors, although the 
potentials
will not be properly Coulombic, but rather involve terms like
$1/(x-a)$, $1/(y-b)$, and $1/(z-c)$.  One can imagine more complicated
variations, such as a spherically symmetric three-dimensional
potential,
say $U(r) = \psi(-r)$, in which case there would be relationships
among
different spherical shells (which would also involve the 
centrifugal
term).

In addition, we can use integrals or derivatives of the digamma
function
to get sectors related by the addition of quantized non-Coulombic
terms.
One interesting example is $U(r) = \psi^\prime(-r)$, as here the extra
term from shifting $r$ by $1$ can be absorbed as a modification to the
centrifugal
term.  As a consequence, shifting from one spherical shell
to another can be reinterpreted
as a modification of the angular momentum term in the radial 
Schr\"odinger
equation to a non-standard value.  Exploration
of this theory is left as an exercise for the interested reader.

\newsec{Extra Dimensions}

The notion of extra dimensions --- extra, that is, beyond the usual
four
of spacetime --- goes back to the 1920s and the work of Kaluza 
\ref\kaluza{Th. Kaluza, Sitzungsber.\ Preuss.\ Akad.\ Wiss.\ Berlin (1921) 966.}
and  Klein \ref\klein{O. Klein, Z. Phys.\ 37 (1926) 895.}.
The Kaluza-Klein approach
has
become one of the standard tools in modern attempts at explaining the 
origins of
the forces, and is especially important in the context of string
theory.

The model we have presented gives a way besides compactification for 
large dimensions
to look small.  Suppose we have a theory in which the fifth dimension 
has the
topology of a ray, and is endowed with a potential $\psi(y)$, where
$y$ 
is the
coordinate along this ray, with the ray stretching along
$y<0$.\foot{Or 
imagine that
a $1/\sin^2(\pi y)$ term has been added, so that we may include
the whole $y$-axis.}
In this extra dimension, which is infinite, the particle will be 
confined to some
sector of, in suitable units, length 1.  Alternatively, one can view 
this as a fifth
dimension which {\it is} of length 1, but for which one has the 
possibility of
different background charges being inserted.  By exchanging the extent
of space for
background charges, we get charge quantization and dimensional 
compactification
automatically,
without invoking the familiar Kaluza-Klein mechanism, and thus without
having to seek a dynamical explanation for the compactification of the
extra dimensions.  Thus what started out as
a new mechanism for generating charge quantization has turned into a
new 
mechanism for
compactifying extra dimensions.

\newsec{Conclusions}

We have shown that a perfectly reasonable quantum mechanical theory in
one
infinite dimension is equivalent to
a quantum theory on a sector of length 1, with the addition of 
background charges,
the values of which must be quantized.  The addition or subtraction of
these
charges arises based on which sector one is considering.
The essential observation is that the amount of charge that can be
added 
or
subtracted is quantized.

We note, too, that this same mechanism helps us see a new way to have 
extra dimensions,
with the price for the apparent compactification of the extra
dimensions 
being the
appearance of quantized background charges.
For both the charge quantization and dimensional reduction
applications, 
it is
clear that one can add any periodic potential to the theory without 
modifying the
results.

It is interesting to see sectors of different charge arising in this
case
without topology.  The similarity to topological charges rests in the 
appearance
of sectors and the quantization of charge.  However, the mechanism is 
entirely
different, and arises here independent of any 
particular
topological considerations.

Transforming this mechanism into an effective and realistic approach
to 
the question
of charge quantization will clearly take some work.  Given the 
importance of the
problem, however, having an alternative way to think about charge 
quantization
(and dimensional reduction/compactification) is clearly of great 
interest, and we
are currently considering ways to enlarge the applicability of this
idea 
to
higher dimensional quantum mechanics and to quantum field theory.
We will leave a more careful consideration of these possibilities to 
future
papers.

DS acknowledges the support of NSF Grant PHY-9970771 under which this 
work was begun.

\listrefs
\bye